\documentclass{nature}
\usepackage{graphicx}
\makeatletter
\let\saved@includegraphics\includegraphics
\AtBeginDocument{\let\includegraphics\saved@includegraphics}
\renewenvironment*{figure}{\@float{figure}}{\end@float}
\makeatother
\usepackage{xcolor}
\usepackage[version=4]{mhchem}
\usepackage{scicite}
\usepackage{mathtools}
\usepackage{times}
\usepackage{amsmath,amssymb,colordvi,mathrsfs,bm,verbatim,dcolumn,epsfig,subfigure,float}
\DeclarePairedDelimiter{\abs}{\lvert}{\rvert}

\usepackage{times}

\title{Observation of the In-plane Anomalous Hall Effect induced by Octupole in Magnetization Space}

\author{Wenzhi Peng$^{1,2 \#}$, Zheng Liu$^{2,3 \#}$, Haolin Pan$^{1,2}$, Peng Wang$^{1,2}$, Yulong Chen$^{1,2}$, Jiachen Zhang$^{1,2}$, Xuhao Yu$^{1,2}$, Jinhui Shen$^{1,2}$, Mingmin Yang$^{1,4}$, Qian Niu$^{2,3\ast}$, Yang Gao$^{2,3\ast}$, Dazhi Hou$^{1,2\ast}$\\
\\
\normalsize{$^{1}$ICQD, School of Emerging Technology, University of Science and Technology of China, Hefei, Anhui 230026, China}\\
\normalsize{$^{2}$Department of Physics, University of Science and Technology of China, Hefei, Anhui 230026, China}\\
\normalsize{$^{3}$CAS Key Laboratory of Strongly-Coupled Quantum Matter Physics, University of Science and Technology of China, Hefei, Anhui 230026, China}\\
\normalsize{$^{4}$Hefei National Laboratory, Hefei, Anhui 230088, China.}\\
\\
\normalsize{$^\ast$Corresponding author. Email: dazhi@ustc.edu.cn(D.H.); ygao87@ustc.edu.cn(Y.G.);}
\\
\normalsize{niuqian@ustc.edu.cn(Q.N.)}\\
\normalsize{\# These authors contributed equally to this work}
}

\begin{document} 


\baselineskip24pt


\maketitle 


\newpage
\begin{abstract}
\
\\

The Anomalous Hall Effect (AHE) manifests as a transverse voltage proportional to magnetization in ferromagnetic materials under the application of a charge current, being an indispensable tool for probing magnetism, especially in nanoscale devices. However, the AHE primarily sensitizes to out-of-plane magnetization, thereby hindering its capacity to discern the in-plane magnetization, a characteristic prevalent in ferromagnetic films. Here we challenge this conventional understanding by demonstrating the in-plane magnetization-induced AHE in iron and nickel, two ubiquitous ferromagnets. This observation of the in-plane AHE is remarkable as it contradicts existing theories that forbid such phenomena in cubic crystal systems. We trace the origin of this unanticipated phenomenon to a hitherto unconsidered octupole of the anomalous Hall conductivity in the magnetization space, a mechanism we propose could enable the detection of in-plane AHE in a wide range of ferromagnetic materials. This work realizes the in-plane AHE in common ferromagnets by exploiting the anomalous Hall conductivity octupole, revealing a new physical origin of the AHE and promising to revolutionize the design of magnetic devices and sensors.

\end{abstract}
\newpage

The Anomalous Hall Effect (AHE) is a phenomenon of paramount importance in condensed matter physics, bridging deep theoretical concepts with practical applications in magnetism sensing\cite{RN00,RN24,RN79,RN78,RN35,RN89,PRA064009,IEEE7386656}. Particularly in nanoscale devices, where conventional magnetometry yields to limitations, the AHE offers unparalleled sensitivity coupled with remarkable ease of integration\cite{Nature428,Nature11switching,CoTc11,LLQSCI,NC2018,Nature19,CrSe2}.However, a significant limitation of the AHE lies in its exclusive detection of out-of-plane magnetization\cite{RN00,RN24}. This inherent bias substantially restricts its application spectrum, especially in ferromagnets where in-plane magnetization is commonly observed\cite{Nature19,TMR,GMR_F,GMR_G,LLQSCI,Feip,Niip,semiip,GaMnAsip}. As a result, detecting in-plane magnetization often necessitates the use of more complex device architectures or relies on indirect methodologies in practical scenarios\cite{GaMnAsip,LLQSCI,Nano16}. Previous theoretical studies have suggested that the in-plane AHE might occur in ferromagnets possessing singular mirror symmetry or strong crystalline anisotropy\cite{yanbinghai,RN76,PRB085123,PRL086802,PRB085411,PRB241103}, and recent experiments demonstrated that Hall effect can be induced by an in-plane magnetic field in low-symmetry nonmagnetic systems\cite{RN82,RN81,RN83,RN80,cui2023inplane}. However, the rarity of materials fulfilling both ferromagnetism and reduced crystalline symmetry poses significant experimental challenges in realizing in-plane AHE in ferromagnets.

In this study, we report the observation of the in-plane AHE unexpected in common ferromagnets, achieved by exploring the multipolar structure of the anomalous Hall conductivity in the magnetization space hitherto neglected in previous researches. The AHE can be characterized by a vector derived from the antisymmetric part of the conductivity tensor, expressed as \(\sigma_{\text{\rm AHE}}^i = \varepsilon_{ijk}\sigma_{jk}\), where \(\varepsilon_{ijk}\) is the Levi-Civita symbol, and the Einstein summation convention applies to repeated indices. Here we represent  $\bm \sigma_{\rm AHE}$ by  the multipoles in the magnetization space where $\bm M=|\bm M| \hat{\bm M}$ with $\hat{\bm M}=(\sin\theta\cos\phi,\sin\theta\sin\phi,\cos\theta)$:
\begin{eqnarray}\label{eq:Sigma}
	\sigma_{\rm AHE}^i=p_{ij}\hat{M}_i+\frac{1}{15}o_{ijkl}\hat{M}_j\hat{M}_k\hat{M}_l,
\end{eqnarray}
in which the first and second terms are the dipole and octupole contributions, respectively. All the even order terms are excluded by the Onsager relation: $\bm{\sigma}_{\text{\rm AHE}}(-\bm{M}) = -\bm{\sigma}_{\text{AHE}}(\bm{M})$, and higher order contributions are omitted here.  The coefficient of the dipole term reduces to a scalar: $p_{ij}=p_0\delta_{ij}$ in ferromagnetic materials with cubic crystal structures, such as Fe and Ni. So the AHE dipole aligns $\bm \sigma_{\rm AHE}$  to $\bm M$ as illustrated in Fig.~1A, which yields an isotropic contribution to the anomalous Hall conductivity and reproduce the familiar orthogonal geometry of the AHE. This indicates that first term in Eq. (\ref{eq:Sigma}) corresponds to $\bm{\sigma}_{\text{AHE}}$ linearly dependent on magnetization which has been the focus of previous works on the AHE, precluding the in-plane AHE in most ferromagnets according to the expression of anomalous Hall current density: \(\bm{J}_{\text{AHE}} = \bm{\sigma}_{\text{AHE}} \times \bm{E}\)\cite{RN24}.

Distinctively, the AHE octupole in the magnetization space, the second term in Eq. (\ref{eq:Sigma}), can be identified as a fundamental origin of the in-plane AHE. $o_{ijk\ell}$ is a rank-4, traceless tensor, symmetric with respect to the last three indices. In cubic crystals, $o_{ijk\ell}$ has only one independent parameters: $o_{xxxx}=o_{yyyy}=o_{zzzz}=-2o_{xxyy}=-2o_{xxzz}=-2o_{yyzz}=-2o_{yyxx}=-2o_{zzxx}=-2o_{zzyy}$. So $\bm \sigma_{\rm AHE}$ can be put into the following form in cubic lattices:
\begin{eqnarray}\label{eq:SigmaDefineCubic}
	\sigma_{\rm AHE}^i=\alpha\hat{M}_i+\beta\hat{M}_i^3,
\end{eqnarray}
where the $x, y~\text{or}~z$ are along the (100), (010), and (001) directions, respectively, $\alpha=p_0-\frac{3}{10}o_{xxxx}$ and $\beta=\frac{1}{2}o_{xxxx}$. In sharp contrast to the dipole term, the octupole term $\beta\hat{M}_i^3$, can cause a misalignment between $\bm \sigma_{\rm AHE}$ and $\bm M$. It is due to the fact that [$\hat{M}_x$,$\hat{M}_y$,$\hat{M}_z$] may not parallel [$\hat{M}^3_x$,$\hat{M}^3_y$,$\hat{M}^3_z$], as illustrated in  Fig.~1B, thus compromising the $\bm \sigma_{\rm AHE} \parallel \bm M$ geometry observed in previous experiments\cite{RN00,RN24,RN55,RN57,RN56}. 
For instance, $\bm M$ in Fe along [103] causes a $\bm \sigma_{\rm AHE}$ component along [1,0,27] through the $\beta\hat{M}_i^3$ term. Such misalignment amounts to a finite anomalous Hall conductivity vector $\bm \sigma_{\text{AHE}}^{\perp}$ normal to $\bm M$, which consequently induces an anomalous Hall voltage at $\bm E$ $\parallel$ $\bm M$, characteristic of the in-plane AHE.




Figure 1C illustrates the experimental setup for measuring the in-plane AHE in Fe(103) films. Before patterned into Hall bar structures, thin films of Fe(103) are epitaxially grown  on MgO(113) substrates using magnetron sputtering, the technical details of which was published elsewhere\cite{RN47}. The Fe(103) orientation is confirmed by the X-ray Diffraction (XRD) and transmission electron microscopy (TEM) patterns as shown in Fig. S1 of the supplementary materials. The crystal lattice of our sample is determined to be 2.860 \AA, close to that of bulk Fe at 2.866 \AA\cite{RN96}. In this setup, when both $\bm M$ and $\bm E$ orient along the in-plane Fe[$30\bar{1}$] direction, $\bm \sigma_{\text{AHE}}^{\perp}$  is expected to induce an in-plane anomalous Hall voltage along the [$0\bar{1}0$] direction.

Figure 1D shows the Hall resistivity ($\rho_{xy}$) as a function of magnetic field $\bm H$ at 300K for $\bm H\parallel$ Fe$[30\bar{1}]$ and $\bm H\parallel$ Fe[$0\bar{1}0$]. The inset of Fig. 1D is the $\bm M$-$\bm H$ curve, indicating that $\bm M$ reaches saturation at  $ \abs{H}<$ 100 Oe for both $\bm H$ orientations. The Hall measurement was carried out by applying charge current along the Fe$[30\bar{1}]$ direction and the Fe film thickness is 100 nm. An evident anomalous Hall signal is observed for $\bm H\parallel$ Fe$[30\bar{1}]$, similar to the $\bm H$ dependence observed in the $\bm M$-$\bm H$ curve. This underscores its origin from the in-plane magnetization, consistent with the in-plane AHE as expected in Fig. 1C. On the other hand, the anomalous Hall signal is absent for $\bm H\parallel$ Fe[$0\bar{1}0$], which can be understood by the vanishing of $\bm \sigma_{\text{AHE}}^{\perp}$ when $\bm \sigma_{\rm AHE}$ $\parallel$  $\bm M$. Collectively, these findings demonstrate the observation of an anomalous Hall effect rooted in the in-plane magnetization of bcc Fe, offering compelling evidence for the in-plane AHE.

Figure 2A depicts an alternative configuration for the Hall measurement, where the orientations of the applied current and the Hall probes are swapped in comparison to the arrangement shown in Fig. 1C. Figure 2B shows the transverse resistivity $\rho_{xy}$ measured with the setup described in Fig. 2A. The magnetic field dependencies of $\rho_{xy}$ for both $\bm H\parallel$ Fe$[30\bar{1}]$ and $\bm H\parallel$ Fe[$0\bar{1}0$] closely resemble those observed in Fig. 1D. This indicates that the Hall signal is predominantly influenced by the magnetization direction rather than the direction of the applied current, which agrees with the $\bm \sigma_{\text{AHE}}^{\perp}$-induced in-plane AHE scenario. 

It is worth noting that if the magnetization is not fully aligned with the magnetic field during the in-plane field sweeping process, its out-of-plane component can also induce an anomalous Hall signal due to the linear $\bm M$ term in the AHE\cite{RN50,RN49}. To discern whether the observed Hall signals in Figs. 1D and 2B are of such a trivial physical origin, we carried out Hall measurement in both setups at magnetic field of 90000 Oe in magnitude, with the result shown in Figs. 2C and 2D. In both figures, the Hall signals remain constant with increasing in-plane magnetic field after saturation, distinct from the attenuation behavior of the AHE induced by the out-of-plane component of tilted magnetization as reported in Ref.\cite{RN50,RN49}. When the magnetic field was aligned normal to the sample, the signal due to the linear AHE can be observed. Combining the measurement of the longitudinal resistivity, the anomalous Hall conductivities of the linear AHE and in-plane AHE are determined to be 1122 $\Omega^{-1} cm^{-1}$ and $-$34.5 $\Omega^{-1} cm^{-1}$, respectively. The opposite signs in the anomalous Hall conductivities of the linear AHE and the  in-plane AHE further indicate their different physical origin. Although the measured in-plane AHE signal in Fe amounts to only a small fraction of the linear AHE signal, its magnitude in $\bm \sigma_{\text{AHE}}$ is still comparable to the largest AHE values found in ambient-temperature 2D ferromagnets\cite{Meng.H_Cr1.2Te2_2021,RN51,RN52}. The in-plane AHE can also be observed in thinner Fe(103) films which shows magnetic field dependence with slightly different magnitudes (please see Fig. S2 in the supplementary materials). 

Figure 3A shows the experimental setup for the Hall measurement in a Ni(111) film in which we also observed the in-plane AHE. Although the Ni(111) film is a well-studied ferromagnetic system in comparison to Fe(103)\cite{H.Ibach_Ni111_1980,A.Meyer_Ni111_2010,Lalmi_PRB_2012}, its the Hall effect  under an in-plane magnetization has not been thoroughly investigated\cite{Ni111}. Figure 3B shows the magnetic field dependence of $\rho_{\rm xy}$ for $\bm H\parallel$ Ni$[10\bar{1}]$ and $\bm H\parallel$ Ni$[2\bar{1}\bar{1}]$. Despite a pronounced planar Hall signal that is symmetric with respect to $\bm{H}$, an antisymmetric component is discernible exclusively when $\bm{H}$ is parallel to Ni$[2\bar{1}\bar{1}]$. Figure 3C shows the field-antisymmetric contribution extracted from Fig. 3B, where only the red curve shows a hysteresis loop with a saturation field close to the value yielded from the $\bm M$-$\bm H$ curve of the Ni(111) film. We can explain the in-plane AHE presence at $\bm H\parallel$ Ni$[2\bar{1}\bar{1}]$ and absence for $\bm H\parallel$ Ni$[10\bar{1}]$ by the $\beta\hat{M}_i^3$ term induced by the AHE octupole in Eq. (\ref{eq:SigmaDefineCubic}) as well. When $\bm{M}$ is parallel to Ni$[10\bar{1}]$, $\bm{\sigma}_{\text{AHE}}$ aligns with Ni$[10\bar{1}]$, maintaining parallelism with $\bm{M}$ and thus lacking an out-of-plane component. Conversely, when $\bm{M}$ is parallel to Ni$[2\bar{1}\bar{1}]$, $\beta\hat{M}_i^3$ generates a $\bm{\sigma}_{\text{AHE}}$ component along Ni$[8\bar{1}\bar{1}]$, resulting in an out-of-plane component $\bm{\sigma}_{\text{AHE}}^{\perp}$ which induced the in-plane AHE. The observation of the in-plane AHE in both iron and nickel suggests its ubiquitousness in ferromagnetic materials.


Figures 4A and 4B depict the experimental setup for measuring the magnetization angle dependence of the Hall signal for Fe(103) and Ni(111) films. The magnitude of $\bm H$ is set at 10000 Oe to ensure that the magnetization is saturated and parallel to $\bm H$ throughout the rotation process. Figure 4C shows the field angle dependence of $\rho_{xy}$ in a 100-nm Fe(103) film, encompassing contributions from both the planar Hall effect (PHE) and the in-plane AHE\cite{RN69,RN68,RN70}. Owing to its even parity in relation to magnetization,  the PHE signal is evident as even harmonics of the field angle $\theta$. Conversely, the in-plane AHE, with its odd parity concerning magnetization, should be characterized by odd harmonics of $\theta$. We employ Fourier analysis to extract the odd and even harmonic constituents of the $\rho_{xy}-\theta$ curve in Fig. 4C, as plotted in Figs. 4D and 4E, respectively. The field angle dependence of the in-plane AHE significantly diverges from the commonly observed $sin(\theta)$ characteristic in the previous AHE studies\cite{RN67,RN65,RN66}. Figure 4F shows the field angle dependence of $\rho_{xy}$ in an 80-nm Ni(111) film. Figure 4G shows the odd harmonic component of the curve in Fig. 4F which manifests a three-fold angle dependence, and Fig. 4H shows a typical planar Hall behavior as in Fig. 4E. We further carried out the same in-plane field rotation measurement of $\rho_{xy}$ in a 100-nm Fe(001) film, and the results are shown in Fig. S3 in the supplementary materials in which the in-plane AHE was not found. These findings reveal qualitative differences in the in-plane AHE behaviors across samples with varying crystalline orientations, which calls for a unified theoretical exploration to elucidate this unconventional behavior not observed in prior AHE researches.

Now we are in a position to derive the magnetization angle dependence of the in-plane AHE from Eq. (\ref{eq:SigmaDefineCubic}), and it is insightful to start from the symmetry analysis of the AHE octupole. Based on Eq.~\eqref{eq:SigmaDefineCubic}, when the magnetization direction varies on a Bloch sphere, the anomalous Hall conductivity can have a nontrivial landscape determined by the crystalline symmetry. With only the dipole term, the AHE is always isotropic in cubic crystals and the Hall deflection plane is always perpendicular to the magnetization. In contrast, the octupole term allows the magnetization lying in the Hall deflection plane to induce a Hall signal as demonstrated in this work. 

In Fe(103) films, when we rotate the magnetization $\bm{M}$ within the (103) plane by an angle $\theta$ with respect to the Fe[010], the corresponding magnetization direction can be expressed as: $\hat{M}_x=-\sin\theta\cos\phi$, $\hat{M}_y=\cos\theta$, and $\hat{M}_z=\sin\theta\sin\phi$, with $\phi$ representing the angle between the (103) and (001) plane. By substituting $\hat{M}_i$ into Eq. (\ref{eq:SigmaDefineCubic}) and isolating the component perpendicular to the (103) plane, we obtain
\begin{eqnarray}\label{eq:103}
	|\bm \sigma_{\text{AHE}}^{\perp}|_{\rm Fe(103)}=\frac{6}{25}\beta_{\rm Fe}\sin^3\theta,
\end{eqnarray}
which yields an excellent match with the AHE data in Fig. 4D using just a single adjustable parameter. Similarly, we obtain the magnetization angle dependence of $|\bm \sigma_{\text{AHE}}^{\perp}|$ in Ni(111) by applying the same derivation:
\begin{eqnarray}\label{eq:111}
	|\bm \sigma_{\text{AHE}}^{\perp}|_{\rm Ni(111)}=\frac{\sqrt{2}}{6}\beta_{\rm Ni}\cos(3\theta),
\end{eqnarray}
which explains the three-fold symmetry of the in-plane AHE data in Fig. 4G. Additionally, the analysis of Fe(001) reveals that \(|\bm \sigma_{\text{AHE}}^{\perp}|_{\text{Fe(001)}}\) vanishes, elucidating the absence of in-plane AHE in Fe(001) films. Thus we reach a comprehensive understanding of the different magnetization angle dependencies of the in-plane AHE in Fe and Ni films, which are essentially different manifestation of the AHE octupole.

Our symmetry analysis indicates that the AHE octuple in the magnetization space is permissible across all Bravais lattices, hinting at its widespread presence in ferromagnetic materials. For instance, in the hexagonal $D_6$ lattice, there exist two nonzero coefficients $o_{ijkl}$, whereas there are three in the tetragonal $D_4$ and $C_{4v}$ and six in the orthorhombic $D_2$ lattices. This increase in the number of nonzero $o_{ijkl}$ coefficients correlates with a more observable in-plane AHE in experiments. Moreover, our analysis identifies several two-dimensional (2D) ferromagnets, including \(\rm CrTe_{2}\), \(\rm Cr_{2}Ge_{2}Te_{6}\), and \(\rm CrSiTe_{3}\) \cite{X.D.Sun_CrTe2_2020,Y.M.Yang_Cr2Ge2Te6_2018,L.D.Casto_CrSiTe3_2015}, also meet the criteria for in-plane AHE due to the influence of the AHE octupole. It is also worthy to note that the in-plane AHE was reported recently in the antiferromagnet $\rm Mn_3Ni_{0.35}Cu_{0.65}N$ and interpreted by the effect of noncollinear magnetic structure\cite{Klaui2024AFMinpAHE}, which might be relevant to the the AHE octuple proposed in this work. So we anticipate the in-plane AHE will emerge as a effective characterization technique for materials with magnetic ordering.

The observation of the in-plane AHE due to the AHE octupole underscores the significance of the distribution of geometric quantities such as the Berry curvature within the magnetization space, highlighting a paradigm shift from the traditional focus on geometric influences in real and momentum spaces\cite{RN89,RN78,Z.Fang_Monopoles_2003}. Based on the concept of the octupole in the magnetization space, additional in-plane Hall-type effects could be discovered. For instance, the anomalous Nernst effect, anomalous Ettinghausen effect, and anomalous thermal Hall effect—each a divergent branch from the AHE\cite{RN59,Uchida_AEE_2018}—it is plausible to expect their in-plane counterparts due to the octupole contribution within suitable ferromagnets. Theoretically, there exists substantial intrigue in discussing the criteria essential for optimizing the octupole contribution and its extended effects across diverse systems. Collectively, these emerging effects due to octupole in the magnetization space may constitute a novel classification of physical phenomena, expanding our understanding and offering new perspectives and paradigms in the study of condensed matter physics.
 
\bibliographystyle{Science}
\bibliography{TAHE}
\
\\
$\textbf{Acknowledgments}$:
This work was supported by the National Key R\&D Program under grant Nos. 2022YFA1403502, the National Natural Science Foundation of China (12234017,12074366). Y. Gao is supported by the  Fundamental Research Funds for the Central Universities (Grant No. WK2340000102). Z. Liu is supported by the National Natural Science Foundation of China (11974327 and 12004369), Fundamental Research Funds for the Central Universities (WK3510000010, WK2030020032), Anhui Initiative in Quantum Information Technologies (No. AHY170000), and Innovation Program for Quantum Science and Technology (2021ZD0302800). The set-up of magnetic sputtering system and thin film growth was assisted by Anhui epitaxy technology co. Ltd. The sample fabrication was supported by the USTC Center for Micro- and Nanoscale Research and Fabrication. The authors thank S. Yang, Z.G. Liang and L.F. Wang for the magnetization measurement. The authors thank B. Wan for the TEM measurement.
\\
$\textbf{Author contributions}$: W.P. and D.H. designed the experiment and analysed the data. W.P., H.P. and M.Y. fabricated the samples and performed characterization of the sample. W.P. carried out the electric measurement. Z.L., D.H., Q.N. and Y.G. developed the explanation of the experiment. W.P., Z.L., P.W. and D.H. wrote the manuscript. Y.C., J.Z., X.Y. and J.S. helped the experiment. All authors discussed the results and commented on the manuscript.
\\
$\textbf{Competing interests}$: The authors declare that they have no competing interests.

\linespread{1.5}

\newcounter{Figure.}
\setcounter{Figure.}{1}
\renewcommand{\thefigure}{\arabic{Figure.}}
\renewcommand{\figurename}{Figure.}
\newpage
\begin{figure}
        \centering
        \includegraphics[width=0.7\linewidth]{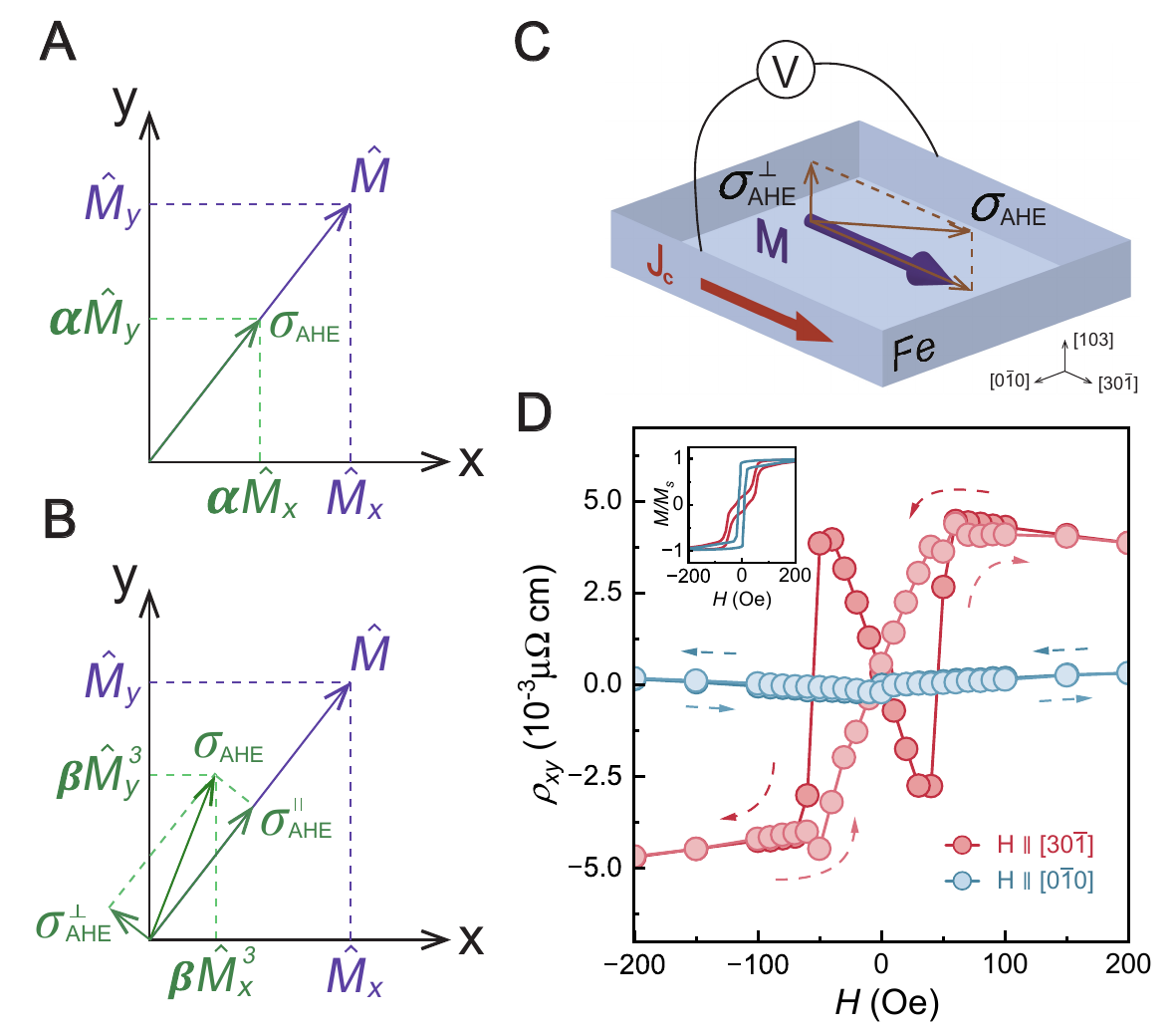}
        \caption{\textbf{Observation of the in-plane anomalous Hall effect in iron.} 
(\textbf{A}) Illustration of the alignment between magnetization \( \bm M \) and the anomalous Hall conductivity vector \(  \bm \sigma_{\text{AHE}}(\bm M) \) with a linear \( \bm M \) response in a cubic lattice, e.g. bcc Fe, where \( \bm \sigma_{\text{AHE}}(\bm M) \) always parallels \( \bm M \).
(\textbf{B}) Illustration of the misalignment between \( \bm M \) and \( \bm \sigma_{\text{AHE}}(\bm M) \) caused by the AHE octupole in a cubic lattice, where \( \bm \sigma_{\text{AHE}}(\bm M) \) can be represented as $(\hat{M}^3_x, \hat{M}^3_y, 0)$. \( \bm \sigma_{\text{AHE}}(\bm M) \) does not parallel \(  \bm M= (\hat{M}_x, \hat{M}_y, 0) \) when \( \hat{M}_x \neq \hat{M}_y \). This misalignment leads to a $\bm \sigma_{\text{AHE}}(\bm M)$ component perpendicular to $\bm M$, denoted as $\bm \sigma_{\text{AHE}}^{\perp}$.
(\textbf{C}) Schematic of the measurement for the in-plane AHE in a bcc Fe(103) film in which \( \bm M \) and current density $\bm J_c$ are both aligned in-plane along Fe$[30\bar{1}]$. \textbf{D}, The in-plane AHE is observed when sweeping the magnetic field \( \bm H \) along the Fe\([ 30\bar{1} ]\) direction (red curve), but is absent when \( \bm H \) is aligned with Fe[$0\bar{1}0$] (blue curve). The inset shows the \( \bm M \)-\( \bm H \) curves for an Fe(103) film with \( \bm H \) sweeping along both Fe\([ 30\bar{1} ]\) (red curve) and Fe[$0\bar{1}0$] (blue curve). Notably, the in-plane AHE signal shows an asymmetric magnetic-field dependence, echoing the saturation behavior of the $\bm M-$$\bm H$ curve.}
        \label{fig1}  
\end{figure}

\setcounter{Figure.}{2}
\begin{figure}
        \centering
        \includegraphics[width=1\linewidth]{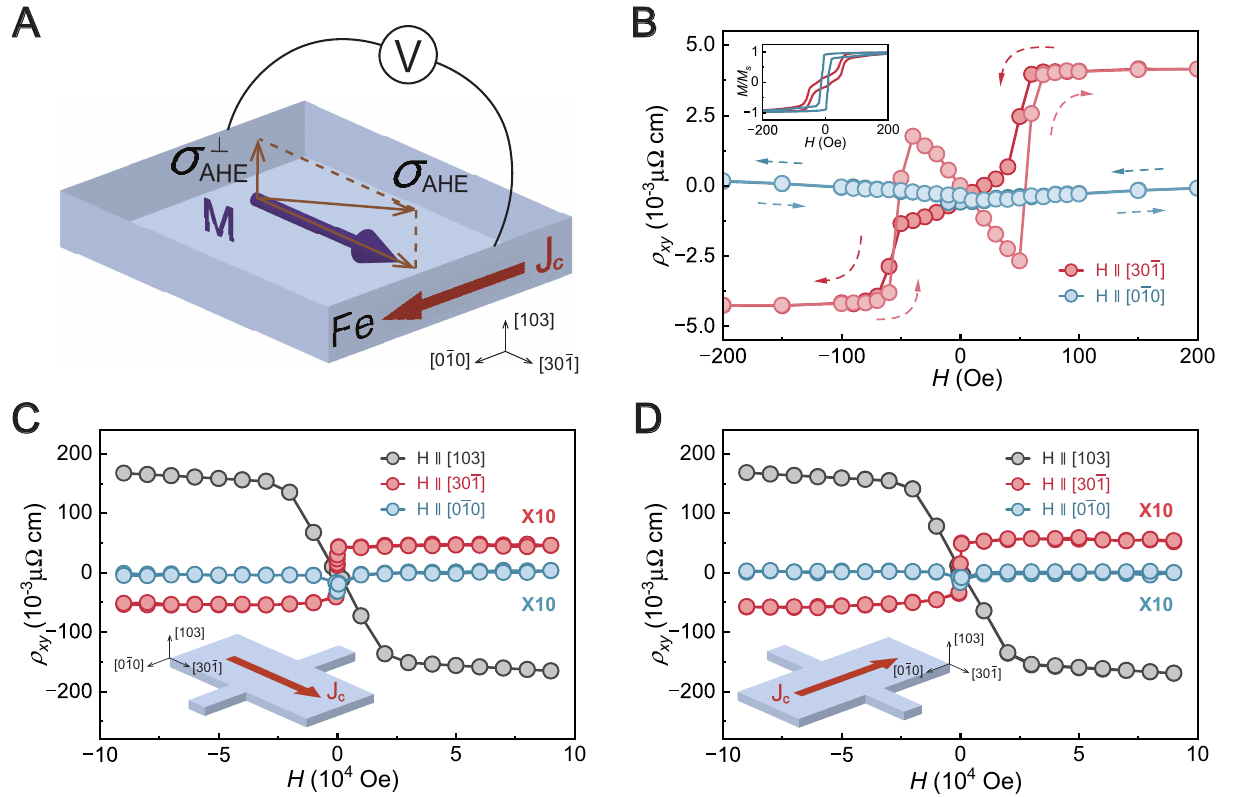}
        \caption{\textbf{Systematic measurement of the in-plane Anomalous Hall effect in Fe(103) films.} 
(\textbf{A}) Illustration of the Hall measurement for an Fe(103) thin film with a current applied along Fe[$0\bar{1}0$].
(\textbf{B}) Hall resistivity measured with $\bm H$ $\parallel$ Fe$[30\bar{1}]$ and $\bm H$ $\parallel$ Fe$[0\bar{1}0]$ for $\abs{H}$ $\leq$ 200 Oe at T = 300 K, and the inset depicts the $\bm M$-$\bm H$ curves measured for the two field orientations. An in-plane AHE signal is observed for \( \bm H \parallel \text{Fe}[30\bar{1}] \) but is absent for \( \bm H \parallel \text{Fe}[0\bar{1}0] \), mirroring the behavior shown in Fig. 1D.
(\textbf{C} and \textbf{D}) Hall resistivity measured at a magnetic field up to 90000 Oe with current applied along Fe$[30\bar{1}]$ and Fe$[0\bar{1}0]$, respectively, where the magnetic field is swept along the perpendicular and two in-plane directions.}
        \label{fig2}  
\end{figure}

\setcounter{Figure.}{3}
\begin{figure}
        \centering
        \includegraphics[width=0.8\linewidth]{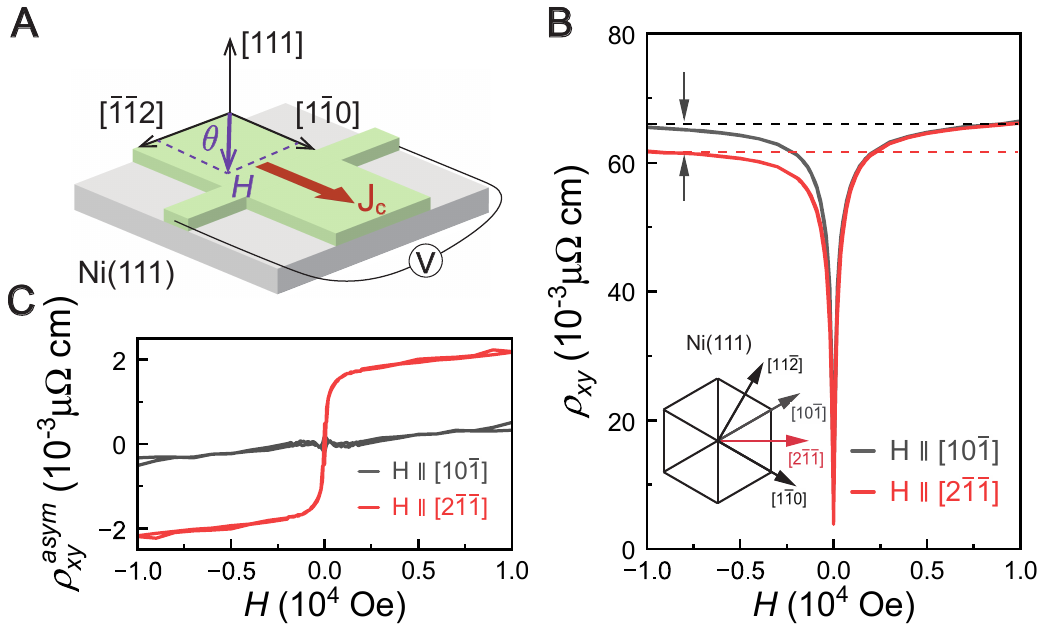}
        \caption{\textbf{Observation of the in-plane anomalous Hall effect in Ni.} 
(\textbf{A}) Illustration of a Hall bar of Ni(111) thin film  with a charge current along Ni$[1\bar{1}0]$.
(\textbf{B}) Hall resistivity measured with $\bm H$ $\parallel$ Ni$[2\bar{1}\bar{1}]$ and $\bm H$ $\parallel$ Ni$[10\bar{1}]$ for $\abs{H}$ $\leq$ 10000 Oe at T = 300 K, and the inset depicts the in-plane crystalline directions of Ni(111). Besides the similar symmetric contributions due to the planar Hall effect\cite{RN69,RN68,RN70}, an antisymmetric contribution is observed for $\bm H$ $\parallel$ Ni$[2\bar{1}\bar{1}]$ but is absent for $\bm H$ $\parallel$ Ni$[10\bar{1}]$.
(\textbf{C}) The field-asymmetric contribution extracted from (\textbf{B}), which reveals that a hysteresis-loop-like Hall signal is only observed for $\bm H$ $\parallel$ Ni$[2\bar{1}\bar{1}]$, consistent with the in-plane AHE induced by the AHE octupole.}
        \label{fig3}  
\end{figure}

\setcounter{Figure.}{4}
\begin{figure}
        \centering
        \includegraphics[width=1\linewidth]{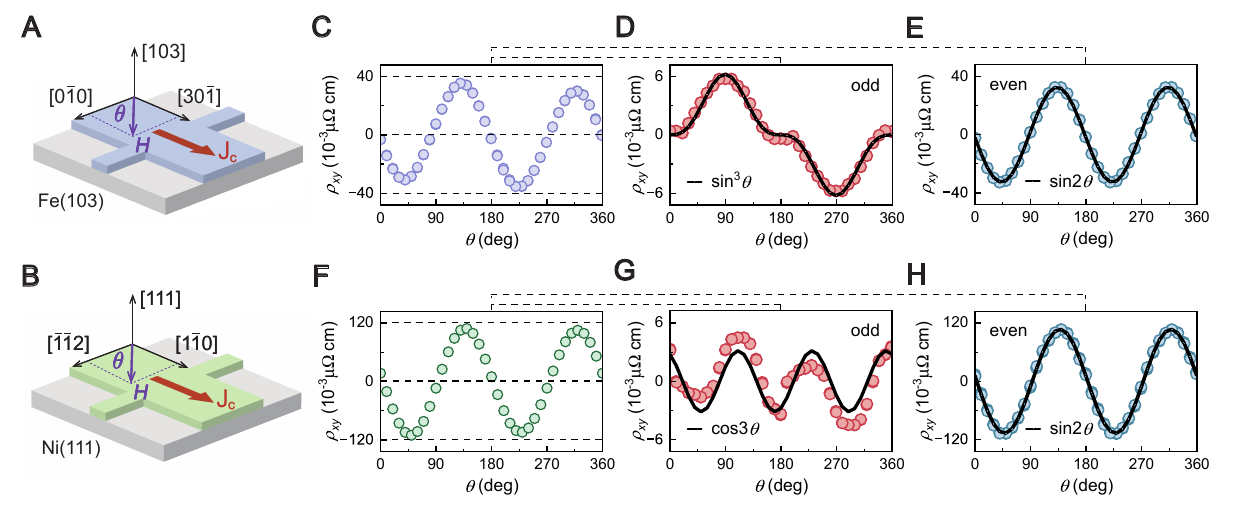}
        \caption{\textbf{Magnetization orientation dependences of the in-plane Anomalous Hall effect.} 
(\textbf{A} and \textbf{B}) Illustrations for the Hall measurement with rotating in-plane magnetic field $\bm H$ for Fe(103) and Ni(111) films, respectively. The magnitude of the magnetic field is set at 10000 Oe, which is sufficiently strong to align $\bm M$ parallel to $\bm H$. 
(\textbf{C}) Field angle dependence of the Hall resistivity in a 100-nm Fe(103) film. 
(\textbf{D} and \textbf{E}) The contributions from odd harmonics and even harmonics derived from (\textbf{C}) respectively. The odd-harmonic pattern agrees well with Eq. (\ref{eq:103}) predicted by the AHE octupole. The even-harmonic contribution  displays a sin(2$\theta$) dependence due to the planar Hall effect. 
(\textbf{F}) Field angle dependence of the Hall resistivity in a 80-nm Ni(111) film. 
(\textbf{G}) The odd-harmonic components in (\textbf{F}), which shows a three-fold field angle dependence with a comparable magnitude as in (\textbf{D}), consistent with Eq. (\ref{eq:111}) predicted by the AHE octupole. 
(\textbf{H}) The even-harmonic components in (\textbf{F}), showing a sin(2$\theta$) dependence due to the planar Hall effect. }
        \label{fig4}  
\end{figure}

\end{document}